\documentclass[aps,reprint,amsmath,amssymb,superscriptaddress]{revtex4-2}

\usepackage{amsmath}    
\usepackage{amssymb}
\usepackage{graphicx}   
\usepackage{verbatim}   
\usepackage[colorlinks,urlcolor=blue,citecolor=blue,linkcolor=blue]{hyperref}
\usepackage{color}      
\usepackage{caption}
\usepackage{subcaption}
\usepackage{hyperref}   
\usepackage{dcolumn}
\usepackage{bm}
\usepackage{epsfig}
\usepackage{epstopdf}
\usepackage{braket}
\usepackage{bbold}

\begin{document}


\title{Dissipative learning of a quantum classifier}

\author{Ufuk~Korkmaz}
\author{Deniz T\"{u}rkpen\c{c}e}
\email[]{dturkpence@itu.edu.tr}

\affiliation{Department of Electrical Engineering, \.{I}stanbul Technical University, 34469 \.{I}stanbul, Turkey}


\date{\today}

\begin{abstract}
The expectation that quantum computation might bring performance advantages in machine learning algorithms motivates the work on the quantum versions of artificial neural networks. In this study, we analyze the learning dynamics of a quantum classifier model that works as an open quantum system which is an alternative to the standard quantum circuit model. According to the obtained results, the model can be successfully trained with a gradient descent (GD) based algorithm. The fact that these optimization processes have been obtained with continuous dynamics, shows promise for the development of a differentiable activation function for the classifier model. 

\end{abstract}


\maketitle

\section{Introduction}\label{intt}

The theory of learning artificial neural networks is foun-ded on mathematical models adapted to the working principle of the human brain, introduced by McCulloch, Pitts and Rosenblatt~\cite{mcculloch_logical_1943,rosenblatt_perceptron_1958}. 
Particularly in the new millennium, in which the computing capacities of computers have increased, it has become a period in which deep learning methods outperform other methods in multi-layer artificial neural networks, which bring many useful applications~\cite{misra_artificial_2010,gu_recent_2018,schmidhuber_deep_2015}.

Quantum computation (QC) brings exciting advantages to computer science and all relevant computational sciences~\cite{bennett_quantum_2000, montanaro_quantum_2016}. 
Although many efforts have been paid for quantum versions of neural networks (QNN), there is no broadly accepted QNN, 
even at a single neuron level~\cite{banchi_quantum_2016,yamamoto_simulation_2018,tacchino_artificial_2019,torrontegui_unitary_2019,mangini_quantum_2021,yan_nonlinear_2020}. 
In addition, quantum noise severely limits the performance of gate-based quantum network proposals.  
Therefore hardware-efficient solutions have began to 
emerge~\cite{pechal_direct_2022,nguyen_evaluation_2022}.

In past work, we proposed a dissipative quantum classifier as a basic unit of QNN hardware, based on repeated interactions protocol~\cite{turkpence_steady_2019,korkmaz_transfer_2022,korkmaz_mimicking_2021,korkmaz_quantum_2022}. Dissipation-based quantum computing has been shown to be equivalent to the standard QC model~\cite{verstraete_quantum_2009}. In the protocol, identical qubit sequences with pure initial quantum state successively interact with a target qubit. The repeated interactions are unitary in the weak coupling limit in a vanishingly small time portion. However, the quantum state of the target qubit is obtained by calculating the reduced dynamics, so that global evolution is a non-unitary process. We dub these identical qubit sequences quantum information reservoir~\cite{deffner_information_2013,deffner_information_driven_2013}. As a result of repeated interactions, the target qubit reaches a steady state in which the diagonal entries of its density matrix become identical to the information reservoir units. This process is known as quantum homogenization~\cite{ziman_diluting_2002}. 

In this task, some amount of information is transferred from the reservoir to the target qubit at the steady state.
This can be interpreted as quantum reservoirs being quantum channels that transfer information to open systems~\cite{blume_kohout_simple_2005,zwolak_redundancy_2017}. All these approaches make sense for open quantum neuron design when the target qubit is connected to more than one information reservoir with arbitrary coupling strengths. In this case, the target qubit reaches a non-trivial steady state depending on the coupling coefficients (weights) and the input data parameters. We have numerically and analytically proposed that this model is an open quantum classifier that returns a binary decision at the steady state when measured by Pauli observables~\cite{turkpence_steady_2019,korkmaz_transfer_2022}. 

In the current work, we study this model in the framework of supervised learning schemes by adopting a gradient descent-based model. To this end, we derive a cost function setting different parameters of the system as variables and examine the availability of the model for learning tasks. We observe that the cost function can be smoothly minimized for all relevant parameters with appropriate differentiability.  

\section{Model and System Dynamics}\label{msd}
\subsection{Classic model}
Binary classification is a subtask for machine learning (ML) covering ANN alongside different models. However, If we discuss, in particular the artificial neural network model, the perceptron is referred to as a basic unit of ANN computing performing binary classification tasks. Technically speaking, a perceptron performs a binary decision $z$ with binary labels $\lbrace 0,1 \rbrace$ depending on the input. In the model, input is formulated as $\varphi_{in}=\bf{x}^T \bf{w}$ where $\bf{x}=[x_1,\ldots x_N]^T$ defines input feature instances and $\bf{w}=[w_1,\ldots w_N]^T$ is the set of corresponding weight vectors. 

The binary output is modulated by, in general a non-linear function $f(.)$ where $z=f(\varphi_{in})$. The decision rule reads $z=0$ if $z=f(\varphi_{in})\geq 0$ and $z=1$ else. The choice of binary labels is arbitrary and can be defined variously depending on the expressivity requirements. Note that, in principle, a perceptron with identity activation can still achieve linear classification. However, non-linear activation functions are desirable for multi-layer ANN learning tasks. Although our model, in principle, is a quantum perceptron with identity activation, we prefer to present our model and related learning tasks as "quantum classifier learning".       

Supervised learning can be defined as a mapping from a feature space to a binary label set 
\begin{equation}\label{Eq:Mapping}
\mathcal{X,Y}\rightarrow \{0,1\}
\end{equation}
where $\mathcal{X}$ and $\mathcal{Y}$ are, respectively, the input and output data of a given a training set 
$\mathcal{S}=(\mathcal{X},\mathcal{Y})$. In this scheme, the $\mathcal{Y}$ part of the training set is the desired output, and the cost function $C$ quantifies how close the actual output is to the desired output.

In analogy with the least squares method, the cost function expression reads
\begin{equation}\label{Eq:Cost1}
C=\frac{1}{2}(\bf{Y}-\bf{A})^2
\end{equation}  
where $\bf{A}$ is the actual and $\bf{Y}$ is the desired output. In general, the weight instances are updated
\begin{equation}\label{Eq:Weight}
\bf{w}_{k+1}=\bf{w}_{k}+\delta\bf{w}_{k}
\end{equation}
iteratively by back-propagation. However, any desired parameter can be adjusted to minimize the cost. 
Among different procedures, we adopt a gradient-descent based method for the training task. In this method, the change
in the parameter reads 
\begin{equation}\label{Eq:delta_w}
\delta {w}_k=-\eta\frac{\partial C}{\partial {w}_k}
\end{equation}
where $\eta$ is a non-negative number, the so-called learning rate, characterizing the speed of the learning task.
As the name of the method implies, the partial derivative expresses the change of the parameter to be adjusted in the direction of the largest descent.  
 
\subsection{Quantum dissipative dynamics}

In this subsection, we discuss the open system dynamics with preliminary definitions. 
As we have pointed out in the previous sections, the model operates by a dissipative protocol.
The input data expressed classically can be rephrased as
\begin{equation}\label{Eq:Weight2}
\varphi_{in}=\bf{x}^T \bf{w}=\sum_i w_i x_i. 
\end{equation}
the weighted summation of the input features. 
In our view, the quantum equivalent of the classic description above reads
\begin{align}\label{Eq:CPTP addition}
\Lambda_t[\varrho_0]=\sum_i P_i\Phi_t^{(i)}{[\varrho_0] }
\end{align}
where $\Phi_t^{(i)}$ is a completely positive trace preserving (CPTP) quantum dynamical map acting on the target qubit $\varrho_0$, $P_i$ is the probability of the map interacting with the $i$th information reservoir. 
The subscript $t$ stands for the time dependence of the maps generated by a physical process
\begin{align}\label{Eq:Unitary}
\Phi^{(i)}_t[\varrho_0]=\text{Tr}_{\mathcal{R}_i}\{U_t(\varrho_0\otimes\varrho_{\mathcal{R}_i})U_t^{\dagger}\}
\end{align}
with a unitary propagator $U_t$ acting on both the target qubit and the reservoir. 
Here, $\rho_{\mathcal{R}_i}$ is the $i$th reservoir quantum state and $\text{Tr}_{\mathcal{R}_i}$ is the partial trace over the $i$th reservoir.  

The quantum reservoirs provide initial quantum data in pure states. Each reservoir is composed of non-correlated, non-interacting two-level quantum systems (subunits) defined by 
\begin{align}\label{Eq:Inf Res}
\rho_{\mathcal{R}_i}=\bigotimes_{k=1}^n\rho_{k}(\theta_i,\phi_i).
\end{align}
the tensor product of finite $n$ subunits. As each subunit is in a pure quantum state, they could initially be prepared by identical Bloch parameters $\rho_{k}(\theta_i,\phi_i)$. This parametrization allows for a dissipative equivalence of the model with parametrized quantum circuits.  

\subsection{Quantum collision model and the quantum \\ classifier}

As mentioned above, the dynamical process of the introduced model relies on a standard quantum collisional model~\cite{scarani_thermalizing_2002,ziman_diluting_2002,nagaj_quantum_2002}. In our proposal, the target qubit undergoes a collisional dissipative process under multiple, independent information reservoirs with arbitrary couplings. In this scheme, the steady state readout of the target qubit by Pauli observables gives the binary classification output. The dynamical process in the presence of the $i$th information reservoir reads
\begin{align}
\Phi^{(i)}_{n\tau}=&\text{Tr}_n \big[ \mathcal{U}_{0 i_n}\ldots\text{Tr}_1[\mathcal{U}_{0 i_1}\left(\varrho_0\otimes\rho_{\mathcal{R}_{i_1}}\right)\mathcal{U}_{0 i_1}^{\dagger}]\otimes\ldots \nonumber \\ 
&\ldots\otimes\rho_{\mathfrak{R}_{i_n}}\mathcal{U}_{0 i_n}^{\dagger} \big].
\end{align}     
Here, $n\tau$ is the time elapsed of the dynamical map for $n$ collisions and $\mathcal{U}_{0 i_k}=\text{exp}[-\text{i}\mathcal{H}^k_{0 i}\tau]$ is the unitary propagator. 
Initially, system plus reservoir quantum states prepared in $ \varrho (0)= \varrho_0 (0)\otimes\rho_{\mathcal{R}_i} $ a tensor product state. Note that the time dependence is only relevant for the target qubit, and after every collision, the reservoir states are reset to their initial state. 

\begin{figure}[!t]
\includegraphics[width=3.3in]{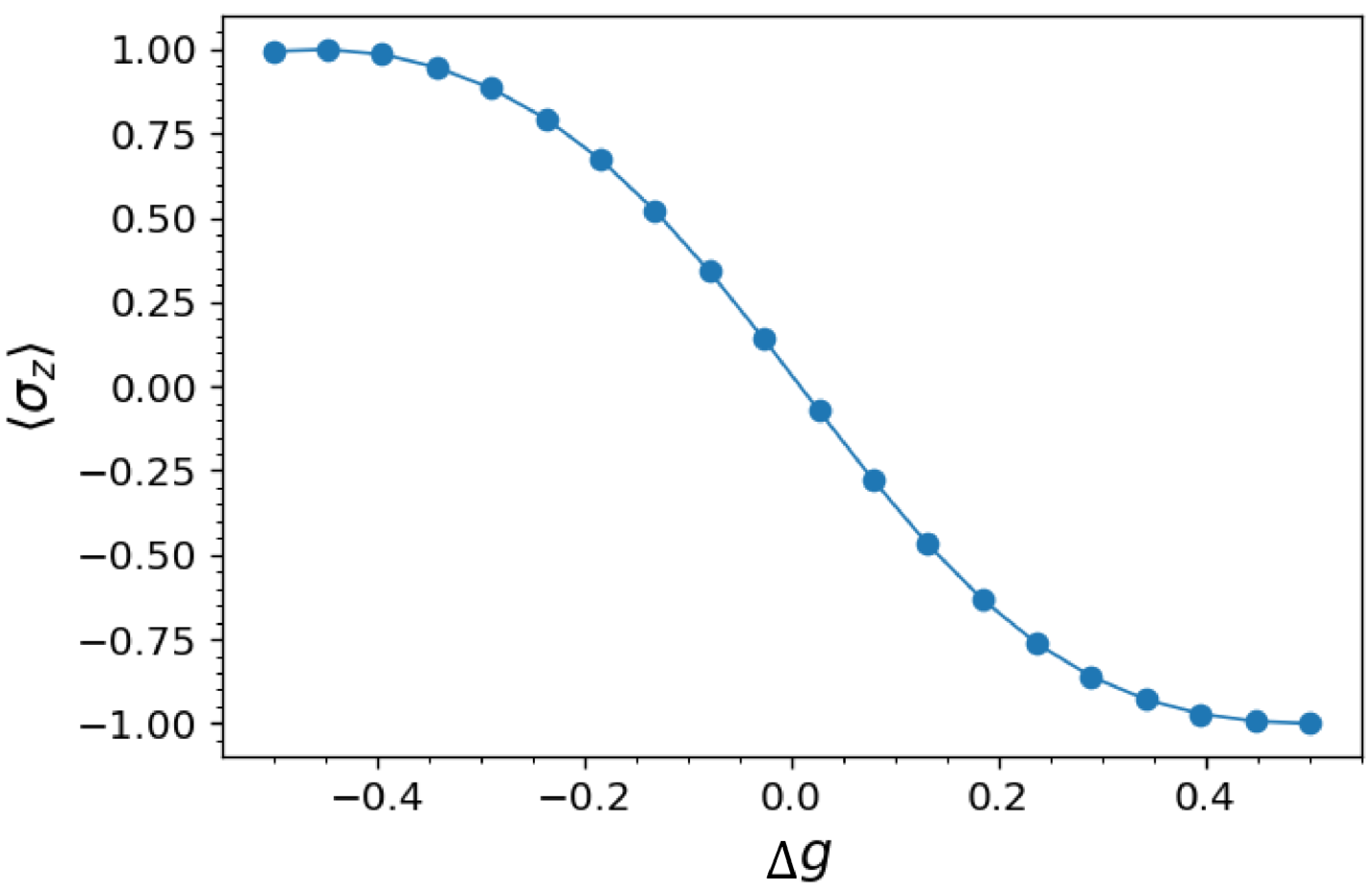}
\caption{ \label{fig:fig1} (Colour online.) The steady state magnetization of the target qubit depending on the variation of the couplings to the reservoirs. Variations of the coupling rates are $g_1=g/2 - \Delta g$, $g_2=g/2 + \Delta g$. Here, $\Delta g$ represents a fraction of $g$ with $g = 0.01$. The probe qubit prepared initially in $\ket{+}=(\ket{e}+\ket{g})/\sqrt{2}$ state and interacted collisionally with the reservoir units $\ket{\Psi(\theta,\phi)}$ with $\theta=0$, $\phi=0$ and $\theta=\pi$, $\phi=0$. The target qubit-reservoir interaction time $\tau=3$ and the coupling coefficient $g$ are dimensionless and scaled by $\omega_r$.} 
\end{figure}

On the other hand, the Hamiltonian governing the system plus reservoir dynamics depicted as $\mathcal{H}= \mathcal{H}_0+\mathcal{H}_{int}$ where 
\begin{align}
\mathcal{H}_0=\frac{\hbar\omega_0}{2}\sigma_0^z+\frac{\hbar\omega_i}{2}\sum_{k=1}^n\sigma_{k_i}^z
\end{align}\label{eq:hfree}
is the free part and 
\begin{equation}
\mathcal{H}_{int}=\hbar\sum_{k=1}^n g_i(\sigma_{0}^+\sigma_{k_i}^- +\text{H.c.}),
\label{eq:hint}
\end{equation}
is the interaction part. Here, respectively, the Pauli-$z$ operator, the Pauli raising and lowering operators read as $\sigma^z$, $\sigma^+$ and $\sigma^-$. The Planck's constant divided by $2\pi$ is taken as $\hbar=1$ throughout the calculations. As a notable point, the value of the coupling coefficient $g_i\ll \omega_0$ ranges within the weak coupling regime where the cross-talk between the reservoirs is avoided. Moreover, the coefficients are proportional to the probabilities $g_i\propto P_i$ in eq.~(\ref{Eq:CPTP addition}) as the quantum equivalent to the weights in the classic model. 

Following the recipe above, the steady state of the target qubit in the presence of $N$ distinct reservoirs reported as the solution of the collisional master equation~\cite{korkmaz_transfer_2022}
\begin{align}\label{Eq:Steadyy} 
\varrho_0^{\text{ss}}=& \frac{1}{\sum_i^N g_i^2 }\sum_{i=1}^N g_i^2\Big( \langle\sigma_{i}^+\sigma_{i}^-\rangle \ket{e}\bra{e}+\langle\sigma_{i}^-\sigma_{i}^+\rangle\ket{g}\bra{g}\nonumber\\
&+i\gamma_1^-\left( \langle\sigma_{i}^+\sigma_{i}^-\rangle-\langle\sigma_{i}^-\sigma_{i}^+\rangle \right)\ket{e}\bra{g}+\text{H.c.}\Big)
\end{align}
where $\ket{e}$ and $\ket{g}$ are the computational basis and $\gamma^-_1=r\tau\sum_{i=1}^N g_i\langle\sigma_{i}^-\rangle$ , $r$ being the interaction rate of the master equation. 
The binary decision at the steady state is read upon the Pauli-$z$ operator acting on the target qubit
\begin{align}\label{Eq:Bloch_ss}
\langle\sigma_{z}^0\rangle^{ss}=\frac{1}{g_{\sum}}\sum_i^N g_i^2\langle\sigma_z\rangle_i
\end{align} 
as the classification identifier where $g_{\sum}=\sum_i g_i^2$. 
Based the eqs~(\ref{Eq:Steadyy}) and (\ref{Eq:Bloch_ss}), finally the binary classification rule reads
\begin{equation}\label{Eq:BinaryCond2}
Decision:
\begin{cases}
0, & \langle \sigma_z^0 \rangle^{ss}=\frac{1}{g_{\sum}}\sum_i^N g_i^2\langle\sigma_z\rangle_i \geq 0        
\\
1, & \text{else}   
\end{cases}
\end{equation}
where  $\langle\sigma_z\rangle_i$ is the $i$th information reservoir magnetization. 
The steady state binary decision expressed by the Pauli-$z$ observable is a summation of the input quantum data 
weighted by respective couplings. This is reasonable as the classic model has a similar expression. 

\begin{figure*}
     \centering
     \begin{subfigure}[t]{0.48\textwidth}
         \centering
         \includegraphics[width=\textwidth]{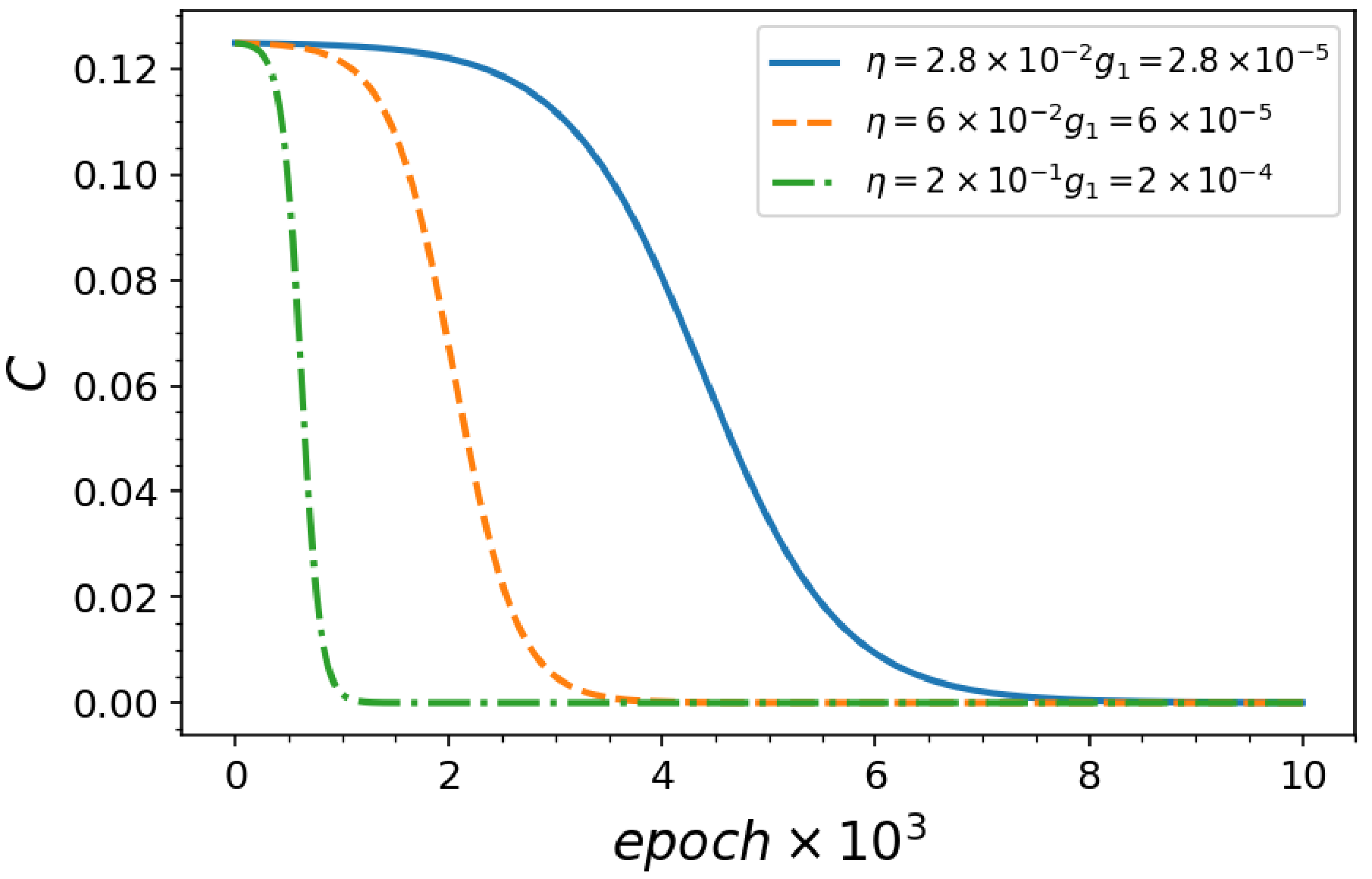}
          \caption{}\label{Fig:Cost_minimize}
     \end{subfigure}
     \hfill
     \begin{subfigure}[t]{0.48\textwidth}
         \centering
         \includegraphics[width=\textwidth]{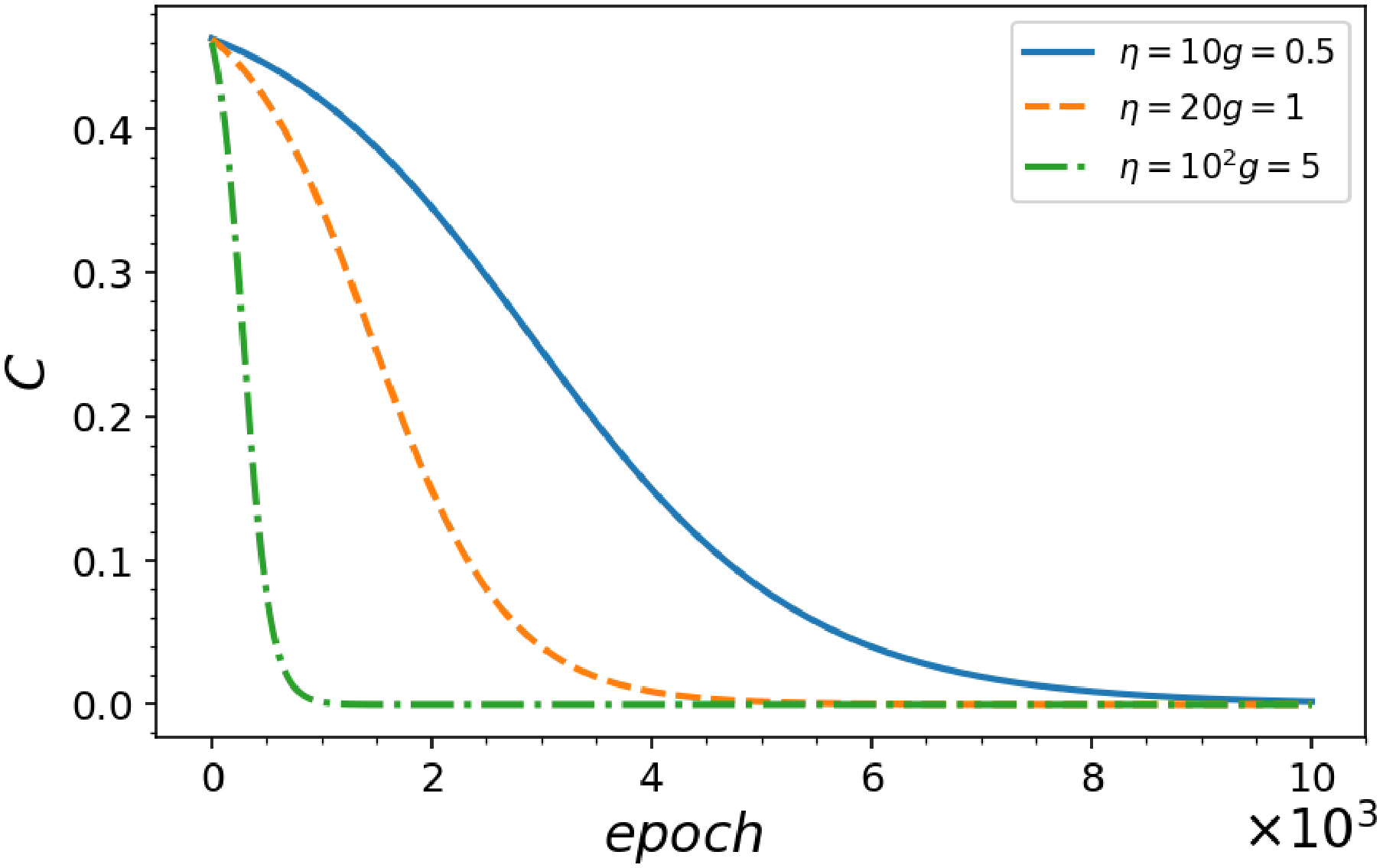}
         \caption{}\label{Fig:Cost_minimize_theta}
     \end{subfigure}
        \caption{Cost function minimization depending on different learning rates against the variation of $g$ and $\theta$ respectively. (a) The initial values of the reservoirs are  $\langle \sigma_z^1\rangle=0.95$, $\langle \sigma_z^2\rangle=-0.15$, $g_1=0.001$, $g_2=0.06$ and $\langle \sigma_z^0\rangle_{des}^{ss}=0.4$, respectively. 
(b) The initial values  $\theta_{1}=170\,^{\circ}$, $\theta_{2}=160\,^{\circ}$, $g=g_1=g_2=0.05$ and $\langle \sigma_z^0\rangle_{des}^{ss}=0$, respectively.}
\end{figure*}

Figure~\ref{fig:fig1} depicts the numerical verification of the introduced model as a benchmark calculation. Here, the target qubit contacted two different information reservoirs with $g_1$ and $g_2$ couplings. The quantum state of the reservoirs are $\ket{\Psi(\theta=0,\phi=0)}\equiv \ket{\uparrow}$ and \\ $\ket{\Psi(\theta=\pi,\phi=0)}\equiv \ket{\downarrow}$, respectively. The dots on the curve represents the steady state magnetization of the target qubit corresponding to $g_1,g_2$ coupling values. These values modulated as $g_1=g/2 - \Delta g$ and $g_2=g/2 + \Delta g$ where $-0.5g<\Delta g<0.5g$. For instance, when $\Delta g=-0.5g$, $g_1=g$ and $g_2=0$ means that the target qubit is in contact only with the first reservoir with the quantum state $\ket{\uparrow}$ and vice versa. In these limits, the steady state magnetization gets $\langle \sigma_z\rangle=1,-1$ and takes intermediate values when $-0.5g<\Delta g<0.5g$ as expected. In the numerical simulation, we have used the realistic parameters of the superconducting circuits in the weak coupling range~\cite{krantz_quantum_2019}. Transmon qubits operate at a resonator frequency $\omega_r\sim 1-10$ GHz with $g\sim 1-100$ MHz effective qubit-qubit coupling.

As depicted above, the relevant parameters are the Bloch parameters $\lbrace\theta,\phi\rbrace$, characterizing the input quantum data. 
Looking more closely at eqs~\ref{Eq:Steadyy} and \ref{Eq:Bloch_ss}, one can see the signatures of input data at the steady state as expected values. The expected values can be related to the Bloch parameters as
\begin{align}\label{Eq:RhoR}
\mathcal{\rho}_{\mathcal{R}_{i}}&=
\begin{bmatrix}
\frac{1+\cos\theta_{i}}{2} & \frac{e^{-i\phi_{i}}}{2}\sin \theta_{i} \\
\frac{e^{i\phi_{i}}}{2}\sin \theta_{i} & \frac{1-\cos\theta_{i}}{2}
\end{bmatrix}\nonumber\\
&:=
\begin{bmatrix}
\langle\sigma_{i}^+\sigma_{i}^-\rangle & \langle\sigma_{i}^-\rangle \\
\langle\sigma_{i}^+\rangle & \langle\sigma_{i}^-\sigma_{i}^+\rangle
\end{bmatrix}
\end{align}
where $\mathcal{\rho}_{\mathcal{R}_{i}}$ is the quantum state of the $i$th reservoir. 
Therefore in our model, Pauli-$z$ and Pauli-$y$ observables can be chosen to extract relevant information 
for $\theta$ and $\phi$ parameters, respectively, at the steady state. Expectation value of the Pauli-$y$ observable 
of the target qubit at the steady state reads 
\begin{align}\label{Eq:Bloch_ss_y}
&\langle\sigma_{y}^0\rangle^{ss}=\frac{-(\gamma_1^- +\gamma_2^+)}{g_{\sum}}\sum_i^N g_i^2\langle\sigma_z\rangle_i
\end{align}  
where $\gamma^-_1=r\tau\sum_{i=1}^N g_i\langle\sigma_{i}^-\rangle$, $\gamma^+_2=r\tau\sum_{i=1}^N g_i\langle\sigma_{i}^+\rangle$. 	
Regarding eqs~(\ref{Eq:Bloch_ss}) and (\ref{Eq:Bloch_ss_y}) together, one evaluates that relevant information of the Bloch parameters can be extracted at the steady state of the target qubit through Pauli observables.

\section{Learning of the model}
In this section, we explore the gradient descent-based learning of the introduced open classifier model.  
First we define the cost function to be optimized as
\begin{equation}\label{Eq:Cost2}
C=\frac{1}{2}(\langle \sigma_{\lambda}^0\rangle_{des}^{ss}-\langle \sigma_{\lambda}^0\rangle_{act}^{ss})^2,
\end{equation}
where $\lambda=\lbrace y,z\rbrace$ denotes the Pauli matrices we choose for specific parameters. 
Here, $\sigma_{\lambda}^0\rangle_{des}^{ss}$ is the desired and $\langle \sigma_{\lambda}^0\rangle_{act}^{ss}$ is the actual steady state expectation values of the target  qubit for the Pauli observable $\sigma_{\lambda}$.  
Definition of the cost function above is similar to \cite{wan_quantum_2017}, however, note that the expected values are obtained in steady states in our task.

Following the classic definitions, we rephrase eqs~(\ref{Eq:delta_w}) and (\ref{Eq:Weight2}) as
\begin{align}
&\bf{\nu}_{k+1}=\bf{\nu}_{k}+\delta\bf{\nu}_{k}\label{Eq:delta_nu}\\
&\delta {\nu}_k=-\eta\frac{\partial C}{\partial {\nu}_k}\label{Eq:delta_C}
\end{align}
where $\nu$=$\lbrace g,\theta,\phi\rbrace$. Therefore, the relevant parameters are the Bloch parameters and the couplings of the target qubit with the reservoirs. We first derive the cost function for $g$ corresponding to the weights in the classic model (see \ref{AppA}). However, we also examine the learning tasks for the Bloch parameters $\theta$ and $\phi$ corresponding to fixed values of $g$.

Figure~\ref{Fig:Cost_minimize} depicts the cost minimization given the parameters against the episodes (the $k$ index) of $\nu=g$ in eqs~(\ref{Eq:delta_nu}) and (\ref{Eq:delta_C}) depending on different values of $\eta$ when the target qubit contacted to two reservoirs. That is, we examine the model using different learning rates (or different optimization speeds). We observe that the optimization always has a smooth feature for different $\eta$s. In the problem, we also observe that the largest possible learning rate is one order of magnitude smaller than the coupling rate.

\begin{figure}
     \centering
     \begin{subfigure}[t!]{0.48\textwidth}
         \centering
         \includegraphics[width=\textwidth]{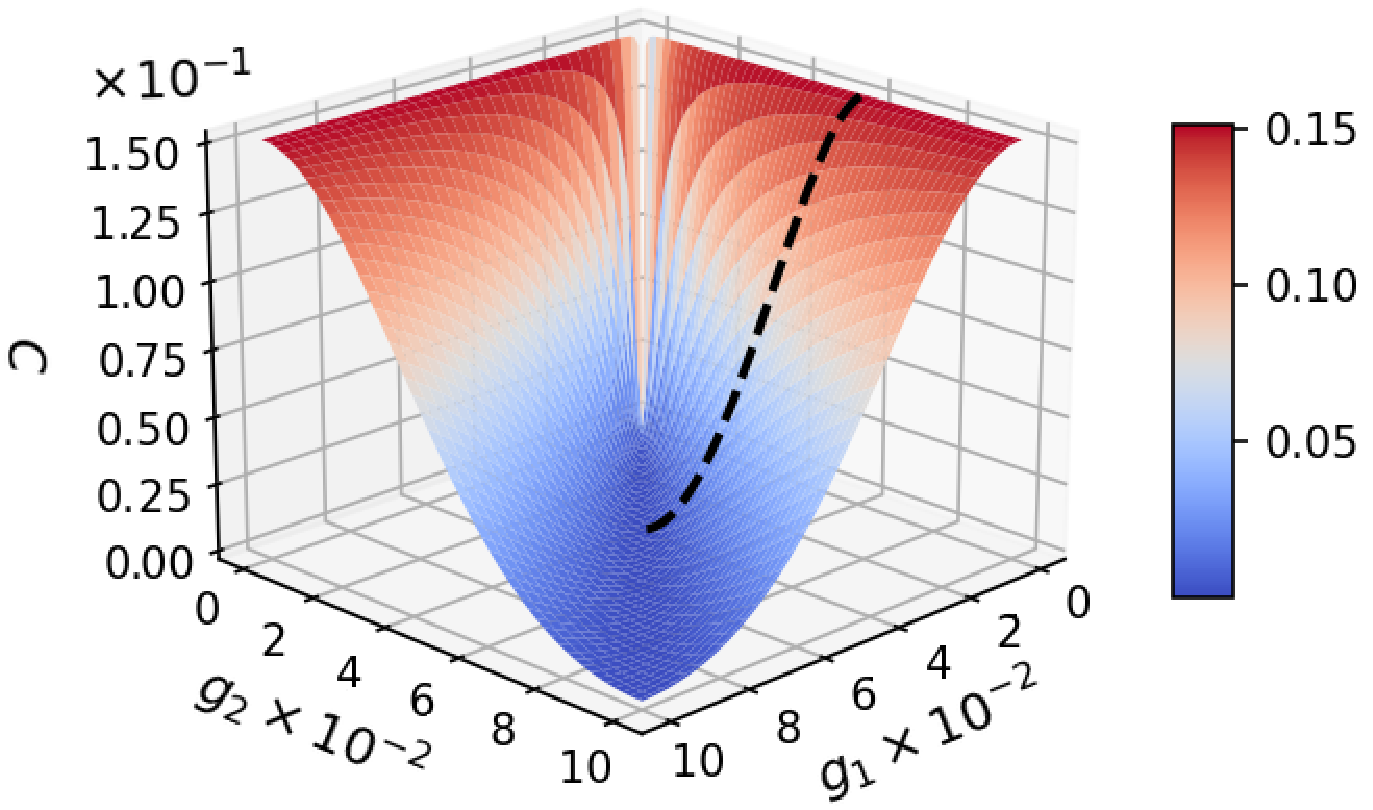}
          \caption{}\label{Fig:Cost_Area1}
     \end{subfigure}
     \hfill
     \begin{subfigure}[t!]{0.48\textwidth}
         \centering
         \includegraphics[width=\textwidth]{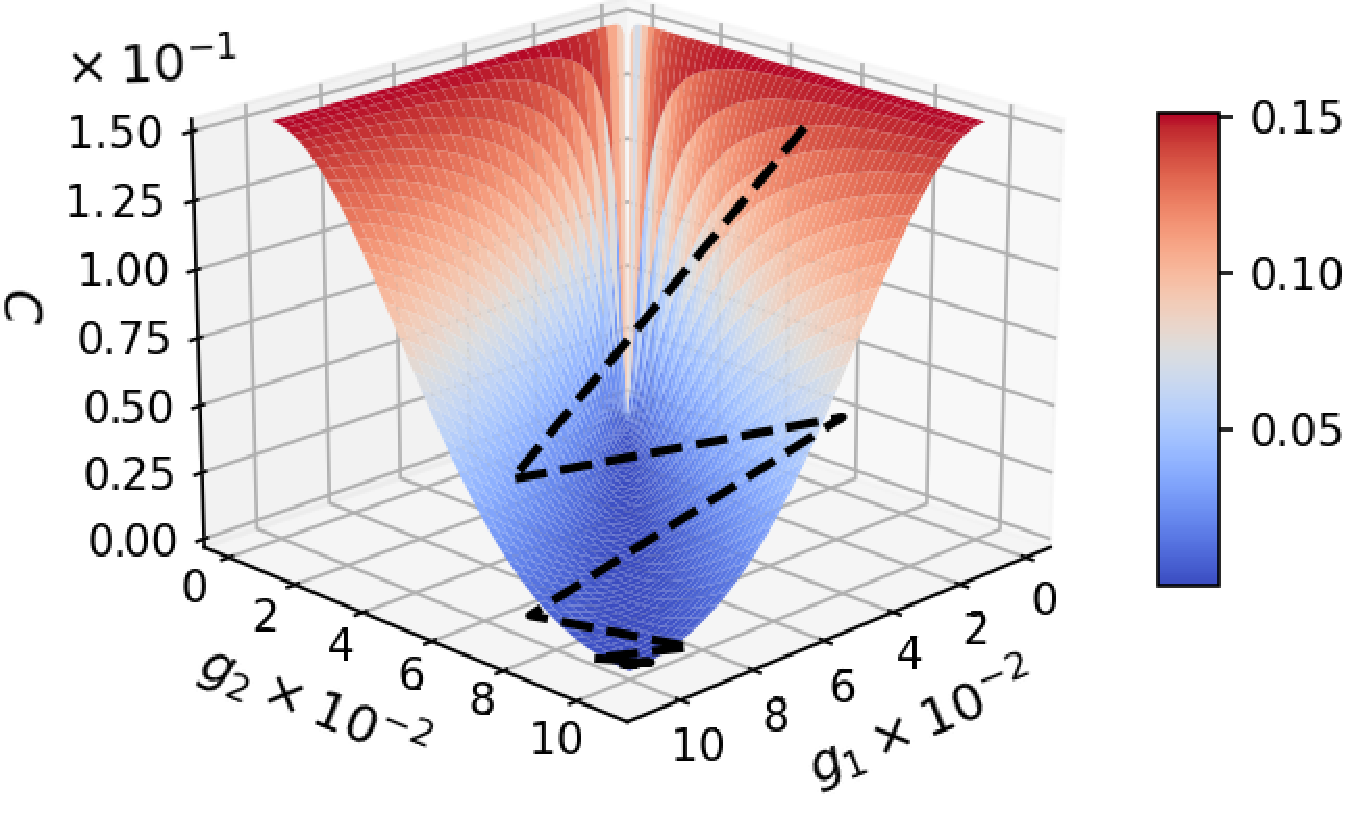}
         \caption{}\label{Fig:Cost_OverShoot}
     \end{subfigure}
        \caption{Cost function minimization with surface depiction against the variation of $g_1$ and $g_2$.
(a) The learning rate $\eta=2.8\times10^{-5}$, $\langle \sigma_z^1\rangle=0.95$, $\langle \sigma_z^2\rangle=-0.15$, 
the target magnetization is $\langle \sigma_z^0\rangle_{des}^{ss}=0.4$ and the initial coupling rates before optimization is $g_1=0.001$, $g_2=0.06$. (b) The learning rate $\eta=2.5\times10^{-2}$, $\langle \sigma_z^1\rangle=0.95$, $\langle \sigma_z^2\rangle=-0.15$, the target magnetization is $\langle \sigma_z^0\rangle_{des}^{ss}=0.4$ and the initial coupling rates before optimization is
$g_1=0.001$, $g_2=0.06$. }
\end{figure}

Figures~\ref{Fig:Cost_Area1} and \ref{Fig:Cost_OverShoot} present the same minimization problem when considering the surface topology of the cost function. In the single target qubit case coupled to two information reservoirs, the structure of the surface cost function seems trivial to optimize without any local plateaus. Therefore, the success of optimization depends on the selection of the learning rate value. In figure~\ref{Fig:Cost_Area1}, the model successfully performs the optimization task with an appropriate learning rate. However, an unstable procedure occurs when a very large value of $\eta$ is selected, as in figure~\ref{Fig:Cost_OverShoot}. Although, in the figure, cost function minimization seems to have been successfully achieved, in most similar problems the iteration value extends beyond the cost function surface. This is known as `overshooting' the minimum.

\begin{figure}
     \centering
     \begin{subfigure}[t!]{0.46\textwidth}
         \centering
         \includegraphics[width=\textwidth]{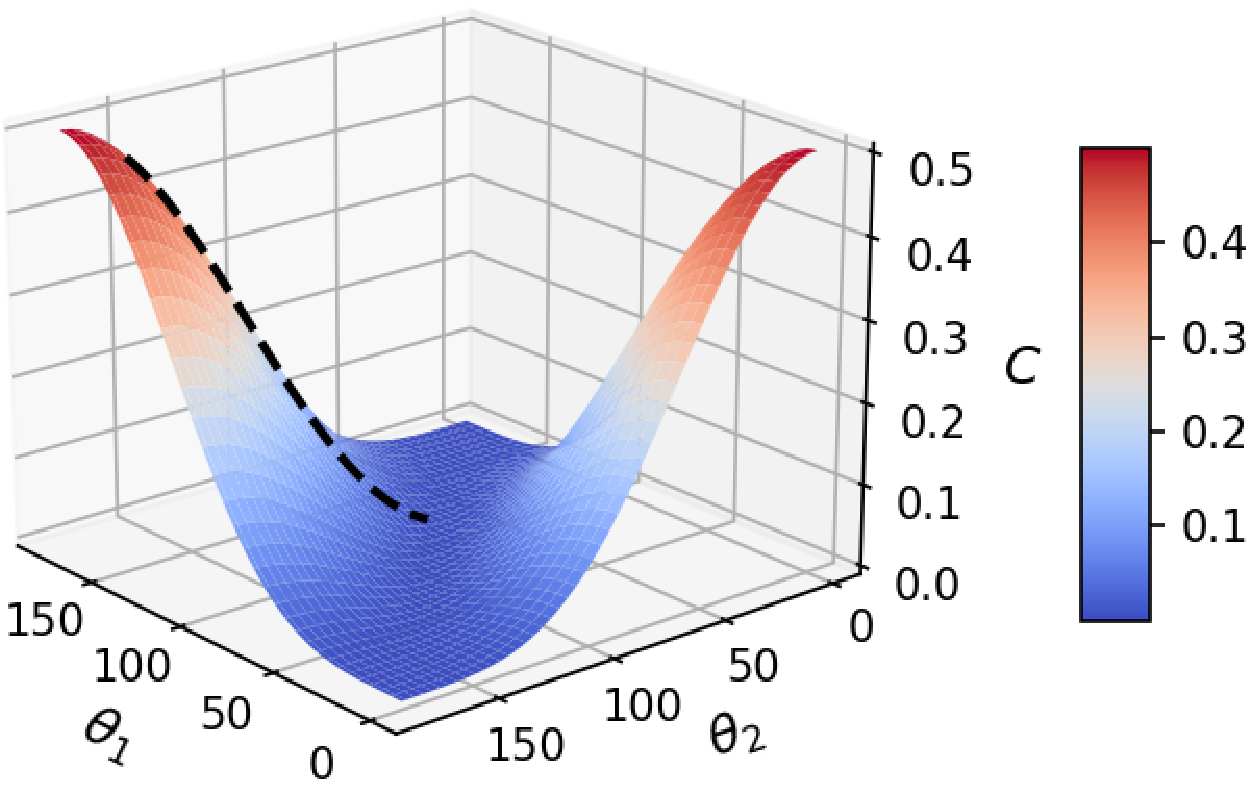}
          \caption{}\label{Fig:Theta_Area}
     \end{subfigure}
     \hfill
     \begin{subfigure}[t!]{0.42\textwidth}
         \centering
         \includegraphics[width=\textwidth]{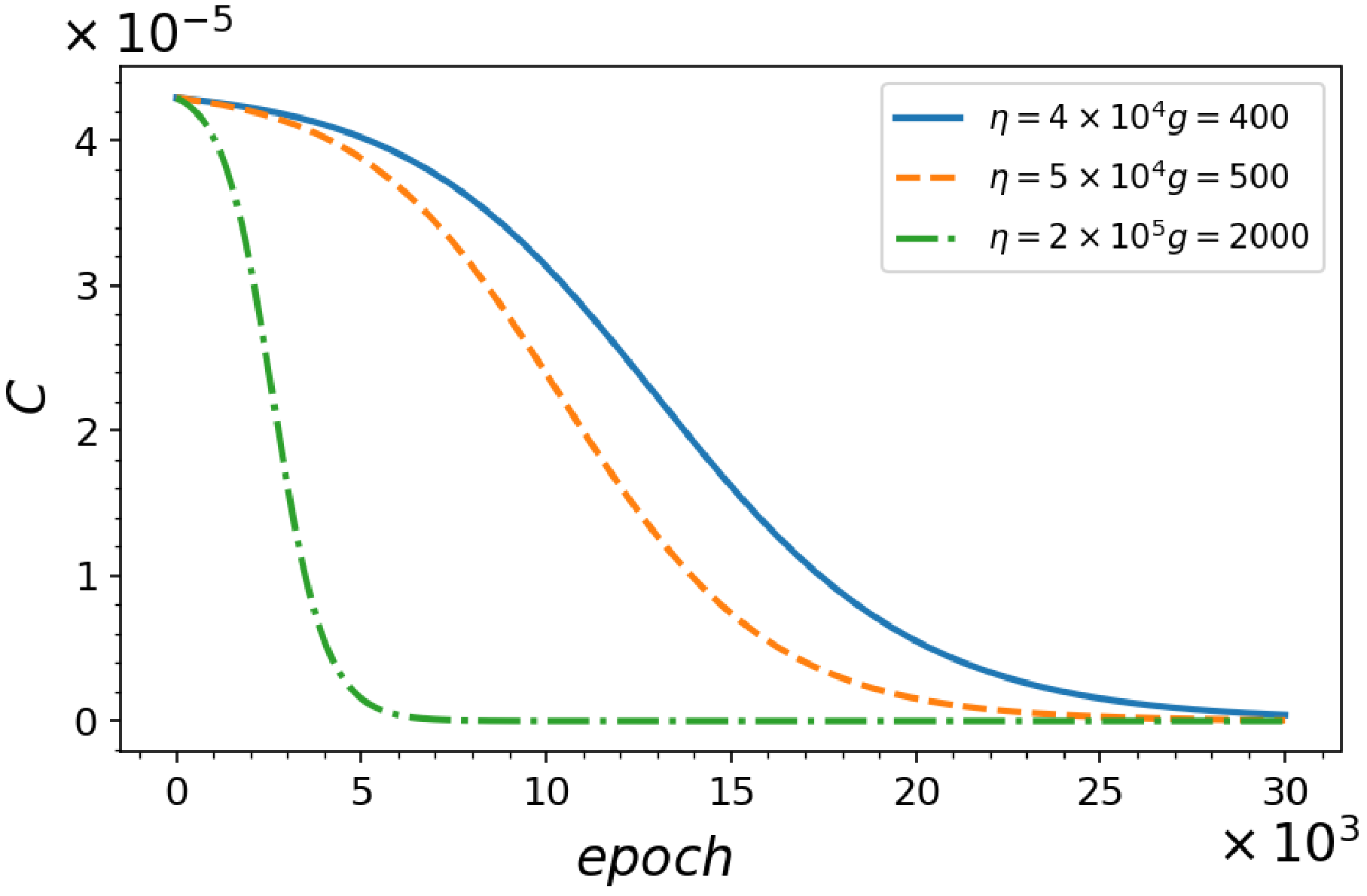}
         \caption{}\label{Fig:Phi_Epoch}
     \end{subfigure}
        \caption{ (a) Cost function minimization with surface depiction against the variation of $\theta_1$ and $\theta_2$. The learning rate $\eta=0.5$, the couplings are $g=g_1=g_2=0.03$ and $\langle\sigma_z^0\rangle_{des}^{ss}=0$.
Initial values for the problem are  $\theta_{1}=170\,^{\circ}$, $\theta_{2}=160\,^{\circ}$. (b) Cost function minimization depending on different learning rates against the variation of $\phi$. Here, $\phi_{1}=350\,^{\circ}$, $\phi_{2}=340\,^{\circ}$,  $\theta_{1}=\theta_{2}=60\,^{\circ}$, $g=g_1=g_2=0.01$ and $\langle \sigma_y^0\rangle_{des}^{ss}=0$, $r=0.36$, $\tau=3$.}
\end{figure}  

Conversely, extremely small learning rate values lead to being stuck in the local minimums. Therefore adaptive tasks, in which the learning rates might take different values during the process, are developed for GD-based methods~\cite{ruder_overview_2017}. We find that for the training of the open classifier model, one order of magnitude around the coupling rate in the weak coupling regime seems a reasonable choice for $\eta$. 

As we have pointed out above, we also examine the cases where the couplings to the reservoirs are fixed. 
In this case, input data parameters are assumed to be adjustable to obtain the desired output. Regarding figure~\ref{Fig:Cost_minimize_theta}, we observe, again, smooth convergence with three orders of magnitude greater learning rate than the coupling coefficient. Corresponding cost function is depicted in figure~\ref{Fig:Theta_Area}. In this case,
the Bloch parameter $\theta$ is iterated to minimize the cost. See eqs~(\ref{Eq:deltac}) and (\ref{Eq:pdg1_g2}) to obtain the cost function in case of the $\theta$ parameter-dependent iteration. The Pauli-$z$ observable is, again, relevant in the calculations. 

Next, we consider the training task concerning the Bloch parameter $\phi$. The Pauli-$y$ observable was chosen to extract $\phi$ parameter-dependent data. This task requires special attention as the proposed classifier operates as an open quantum system, driven by non-equilibrium reservoirs. Steady states bear mixed quantum states in which quantum coherent information is irreversibly lost. However, some non-vanishing quantum coherence may exist when the system is driven by non-equilibrium environments.~\cite{scully_extracting_2003,karevski_quantum_2009}. In our case, eq.~(\ref{Eq:Steadyy}) demonstrates that the target qubit retains quantum coherence at the steady state as the non-diagonal part of the density matrix is non-zero. In addition, eq.~(\ref{Eq:Bloch_ss_y}) states that the steady coherence is weighted by the coupling coefficients where it can be parametrized by $\phi$ through the Pauli-$y$ observable. 

\begin{figure}[!t]
\includegraphics[width=3.0in]{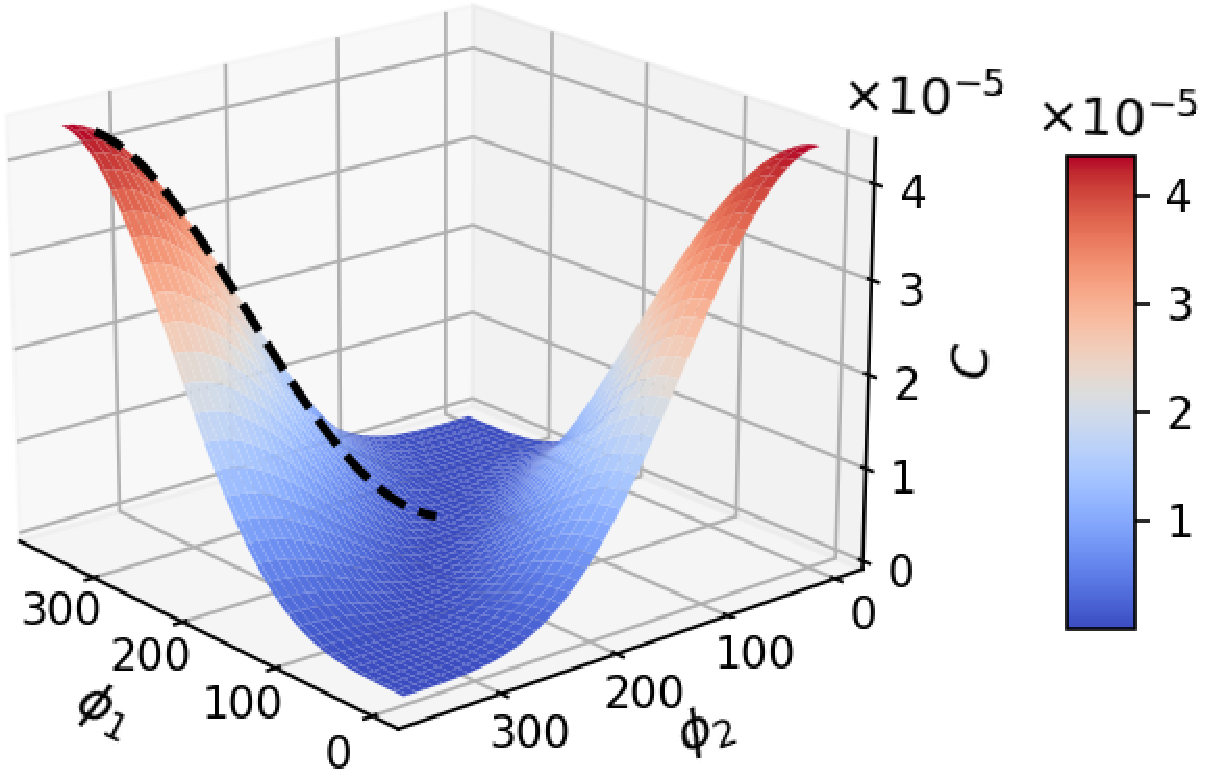}
\caption{ \label{fig:fig7} (Colour online.) Cost function minimization with surface depiction against the variation of $\phi_1$ and $\phi_2$. Here, the initial values for the dashed (optimization) line are $\eta=500$, $\phi_{1}=350\,^{\circ}$, $\phi_{2}=340\,^{\circ}$,  $\theta_{1}=\theta_{2}=60\,^{\circ}$, $g=g_1=g_2=0.01$ and $\langle \sigma_y^0\rangle_{des}^{ss}=0$, $r=0.36$, $\tau=3$ respectively.} 
\end{figure}

Figure~\ref{Fig:Phi_Epoch} shows the cost minimization depending on different learning rates. 
The cost function values around the $\times 10^{-5}$ scale reveals a small coherence value at the steady state compared to the diagonal elements of the target qubit density matrix. In addition, the $\eta$ value for the $\phi$ parameter optimization has the largest value compared to the optimizations for the $g,\theta$ parameters. Finally, figure~\ref{fig:fig7} represents he cost function minimization considering the update of Bloch parameters $\phi_1$ and $\phi_2$. The 3D surface of the cost function is similar to figure~\ref{Fig:Theta_Area}, only differing by the value 
of the learning rate.  

If a comment is made by evaluating all the results together, we see that the proposed classifier is suitable for GD-based training schemes. Moreover, open system dynamics allows for smooth convergence in learning tasks which makes the model favourable for multi layer feed-forward networks once an activation function is introduced. Since binary classification is a task in itself for ML, the model we propose is a candidate to be a trainable model in ML processes, even when considered alone.  

\section{Conclusions}

In this study, we examined the training of a classifier model based on the open quantum model in different parameter spaces with the GD-based method. Using our analytical results, we have derived cost functions for three different parameters for training our model and made calculations that minimize the cost functions with the gradient descent algorithm. Obtaining the classification response of the model in a stationary state makes the system dynamics continuous dynamics. As a result of this, we achieved optimization of the model, namely its training with smooth, continuous results. Since the training processes are continuous, which means that they are differentiable, it is concluded that the model we propose is suitable for developing an activation function and using it in larger quantum networks. In addition, although the classification result is taken in a stationary state, it becomes possible to train in all Bloch parameter spaces as well as the coupling coefficients by the steady quantum coherence.

Our study revealed that the derived cost functions are trained at different values of learning rates for corresponding parameters. In our model, cost functions successfully minimized with appropriate learning coefficients.

\section*{Acknowledgment}

The authors acknowledge support from the Scientific and Technological Research Council of Turkey (TÜBİTAK-Grant No. 120F353). The authors also wish to extend special thanks to the  Cognitive Systems Lab in the Department of Electrical Engineering providing the atmosphere for motivational and stimulating discussions.

\appendix
\section{Derivation of the cost function}\label{AppA} 

In this section, we present the mathematical justifications for numerical calculations in the text.
First, we substitute $\nu=g$ in eq.~(\ref{Eq:delta_C}) 
\begin{equation}\label{Eq:deltag}
\delta {g}_i=-\eta\frac{\partial C}{\partial {g}_i}.
\end{equation}
and obtain the cost function expression taking the partial derivative with respect to the coupling constant $g$.
\begin{equation}\label{Eq:deltac}
\frac{\partial C}{\partial {g}_i}=(\langle \sigma_z^0\rangle_{des}^{ss}-\langle \sigma_z^0\rangle_{act}^{ss})(-\frac{\partial \langle \sigma_z^0\rangle_{act}^{ss}}{\partial {g}_i})
\end{equation}

In our current example, we have two information reservoirs corresponding to specific magnetizations.  
Therefore, the actual steady state magnetization (eq.~(\ref{Eq:Bloch_ss})) reads as
\begin{equation}\label{Eq:Actual}
A=\langle \sigma_z^0\rangle_{act}^{ss}=\frac{g_{1}^{2}\langle \sigma_z^1\rangle+g_{2}^{2}\langle \sigma_z^2\rangle}{g_{1}^{2}+g_{2}^{2}}.
\end{equation}
According to the recipe to derive the cost function, the partial derivatives with respect to $g_1$ and $g_2$ separately obtained as
\begin{align}\label{Eq:pdg1_g2}
&\frac{\partial A}{\partial {g}_1}=\frac{2g_{1}\langle \sigma_z^1\rangle (g_{1}^{2}+g_{2}^{2})-2g_{1}(g_{1}^{2}\langle \sigma_z^1\rangle+g_{2}^{2}\langle \sigma_z^2\rangle)}{(g_{1}^{2}+g_{2}^{2})^{2}}\nonumber\\
&\frac{\partial A}{\partial {g}_2}=\frac{2g_{2}\langle \sigma_z^2\rangle (g_{1}^{2}+g_{2}^{2})-2g_{2}(g_{1}^{2}\langle \sigma_z^1\rangle+g_{2}^{2}\langle \sigma_z^2\rangle)}{(g_{1}^{2}+g_{2}^{2})^{2}}
\end{align}
In our example, the desired magnetization is $\langle \sigma_z^0\rangle_{des}^{ss}=0.4$ a constant value in the cost function. After substituting eqs.~(\ref{Eq:Actual}) and~(\ref{Eq:pdg1_g2}) in eq.~(\ref{Eq:deltac}), the expression obtained after substituting them in eq.~(\ref{Eq:deltag}), eq.~(\ref{Eq:delta_nu}) becomes as follows:
\begin{align}\label{Eq:training_1}
&(g_1)_{k+1}=(g_1)_{k}+\delta(g_1)_{k}\nonumber\\
&(g_2)_{k+1}=(g_2)_{k}+\delta(g_2)_{k}.
\end{align}

Next,  we substitute $\nu=\theta$ in eq.~(\ref{Eq:delta_C}) as
\begin{equation}\label{Eq:delta_theta}
\delta {\theta}_i=-\eta\frac{\partial C}{\partial {\theta}_i}.
\end{equation}
Regarding eq.~(\ref{Eq:RhoR}), one can easily see that the magnetization of the $i$th reservoir is 
$\langle\sigma_z\rangle_i=\langle\sigma_{i}^+\sigma_{i}^-\rangle - \langle\sigma_{i}^-\sigma_{i}^+\rangle$. 
Therefore, azimuth parameter-dependent expression of the magnetization can be easily written as  $\langle\sigma_z\rangle_i=\cos\theta_{i}$.

Equation~(\ref{Eq:deltac_theta}) is obtained when we take the partial derivative of the cost function with respect to $\theta$.
\begin{equation}\label{Eq:deltac_theta}
\frac{\partial C}{\partial {\theta}_i}=(\langle \sigma_z^0\rangle_{des}^{ss}-\langle \sigma_z^0\rangle_{act}^{ss})(-\frac{\partial \langle \sigma_z^0\rangle_{act}^{ss}}{\partial {\theta}_i})
\end{equation}

In our current example, we have two information reservoirs corresponding to specific magnetizations.  
Therefore, the actual steady state magnetization (eq.~(\ref{Eq:Bloch_ss})) reads as
\begin{equation}\label{Eq:Actual_theta}
A=\langle \sigma_z^0\rangle_{act}^{ss}=\frac{g_{1}^{2}\cos\theta_{1}+g_{2}^{2}\cos\theta_{2}}{g_{1}^{2}+g_{2}^{2}}.
\end{equation}

According to the recipe to derive the cost function, the partial derivatives with respect to $\theta_{1}$ and $\theta_{2}$ separately obtained as
\begin{align}\label{Eq:pdt1_t2}
&\frac{\partial A}{\partial {\theta}_1}=-\frac{g_{1}^{2}\sin\theta_{1}}{g_{1}^{2}+g_{2}^{2}}\nonumber\\
&\frac{\partial A}{\partial {\theta}_2}=-\frac{g_{2}^{2}\sin\theta_{2}}{g_{1}^{2}+g_{2}^{2}}
\end{align}
In our example, the desired magnetization is $\langle \sigma_z^0\rangle_{des}^{ss}=0$ a constant value in the cost function. After substituting eqs~(\ref{Eq:Actual_theta}) and~(\ref{Eq:pdt1_t2}) in eq.~(\ref{Eq:deltac_theta}), the expression obtained after substituting them in eq.~(\ref{Eq:delta_theta}), eq.~(\ref{Eq:delta_nu}) becomes as follows:
\begin{align}\label{Eq:training_theta}
&\bf{(\theta)}_{k+1}=\bf{(\theta_1)}_{k}+\delta\bf{(\theta_1)}_{k}\nonumber\\
&\bf{(\theta_2)}_{k+1}=\bf{(\theta_2)}_{k}+\delta\bf{(\theta_2)}_{k}.
\end{align}

Let's edit Eq.~(\ref{Eq:delta_nu}) for $\nu=\phi$

\begin{equation}\label{Eq:delta_phi}
\delta {\phi}_i=-\eta\frac{\partial C}{\partial {\phi}_i}.
\end{equation}

Equation~(\ref{Eq:deltac_phi}) is obtained when we take the partial derivative of the cost function with respect to $\phi$.
\begin{equation}\label{Eq:deltac_phi}
\frac{\partial C}{\partial {\phi}_i}=(\langle \sigma_y^0\rangle_{des}^{ss}-\langle \sigma_y^0\rangle_{act}^{ss})(-\frac{\partial \langle \sigma_y^0\rangle_{act}^{ss}}{\partial {\phi}_i})
\end{equation}
In our current example, we have two information reservoirs corresponding to specific magnetizations.  
Therefore, the actual steady state magnetization (eq.~(\ref{Eq:Bloch_ss})) by using eq.~(\ref{Eq:RhoR})  reads as
\begin{widetext}
\begin{align}\label{Eq:Actual_phi}
A=\langle \sigma_y^0\rangle_{act}^{ss}=-r\tau\frac{g_{1}^{3}\sin\theta_{1}\cos\theta_{1}\cos\phi_{1}+g_{1}g_{2}^{2}\sin\theta_{1}\cos\theta_{2}\cos\phi_{1}+g_{1}^{2}g_{2}\cos\theta_{1}\sin\theta_{2}\cos\phi_{2}+g_{2}^{3}\sin\theta_{2}\cos\theta_{2}\cos\phi_{2}}{g_{1}^{2}+g_{2}^{2}}.
\end{align}
\end{widetext} 
According to the recipe to derive the cost function, the partial derivatives with respect to $\phi_{1}$ and $\phi_{2}$ separately obtained as
\begin{align}\label{Eq:pdp1_p2}
&\frac{\partial A}{\partial {\phi}_1}=r\tau\frac{g_{1}^{3}\sin\theta_{1}\cos\theta_{1}\sin\phi_{1}+g_{1}g_{2}^{2}\sin\theta_{1}\cos\theta_{2}\sin\phi_{1}}{g_{1}^{2}+g_{2}^{2}}\nonumber\\
&\frac{\partial A}{\partial {\phi}_2}=r\tau\frac{g_{1}^{2}g_{2}\cos\theta_{1}\sin\theta_{2}\sin\phi_{2}+g_{2}^{3}\sin\theta_{2}\cos\theta_{2}\sin\phi_{2}}{g_{1}^{2}+g_{2}^{2}}
\end{align}
In our example, the desired magnetization is $\langle \sigma_y^0\rangle_{des}^{ss}=0$ a constant value in the cost function. After substituting eqs~(\ref{Eq:Actual_phi}) and~(\ref{Eq:pdp1_p2}) in eq.~(\ref{Eq:deltac_phi}), the expression obtained after substituting them in eq.~(\ref{Eq:delta_phi}), eq.~(\ref{Eq:training_phi}) becomes as follows:
\begin{align}\label{Eq:training_phi}
&\bf{(\phi)}_{k+1}=\bf{(\phi_1)}_{k}+\delta\bf{(\phi_1)}_{k}\nonumber\\
&\bf{(\phi_2)}_{k+1}=\bf{(\phi_2)}_{k}+\delta\bf{(\phi_2)}_{k}.
\end{align}

\end{document}